\documentclass[twocolumn,showpacs,preprintnumbers,amsmath,amssymb]{revtex4}

\usepackage{graphicx}
\usepackage{dcolumn}
\usepackage{bm}

\newcommand{\di}{\displaystyle}
\renewcommand{\dh}{\hspace{-2 mm}\di}

\begin{document}

\title{Long-range attraction between particles
in dusty plasma\\ and partial surface
tension of dusty phase boundary}

\author{A. G. Bashkirov}
\email{abas@idg.chph.ras.ru}

\affiliation{Institute Geospheres Dynamics of Russian Academy of
Sciences, Moscow, Russia} 

\date{\today}

\begin{abstract}
Effective potential of a charged dusty particle moving in
homogeneous plasma has a negative part that provides attraction
between similarly charged dusty particles. A depth of this
potential well is great enough to ensure both stability of crystal
structure of dusty plasma and sizable value of surface tension of
a boundary surface of dusty region. The latter depends on the
orientation of the surface relative to the counter-ion flow,
namely, it is maximal and positive for the surface normal to the
flow and minimal and negative for the surface along the flow. For
the most cases of dusty plasma in a gas discharge, a value of the
first of them is more than sufficient to ensure stability of
lenticular dusty phase void oriented across the counter-ion flow.
\end{abstract}


\maketitle
\section{Introduction}

A gas--solid phase transition observed in different dusty
laboratory plasmas \cite{1,2,3,4,6,8,9,10}. It counts in favor of
presence of strong long-range attraction between similarly charged
dusty particles. By now some physical mechanisms are proposed to
explain formation of a regular arrangement of micron-sized
particles embedded in a gas discharged plasma. A part of them is
based on the account for electrostatic fields of strata and walls
of discharge tubes\cite{17}. In doing, so a generality of solution
of the problem of the effective attractive potential is lost and,
in particular, the situation of dusty plasma crystal in a thermal
dusty plasma of gas burner \cite{6} where there are no walls and
external fields drops out of such interpretations.

The most promising approach holds that the ion streaming motion
causes an attractive wake potential behind the dust particles.
Originally it was developed \cite{20,21} for particular case of
supersonic flows which is realized in the sheath of radiofrequency
discharges. Later, it was extended with participation of one of
the authors of Ref. \cite{20} to the case of subsonic ion flows
\cite{vlad}, but physics of shielding of dust charged particles
was supposed to be strongly modified with mandatory regard to
anisotropy and asymmetry of the ion temperature in the sheath.

Below is shown that the attraction between similarly charged
particles can be resulted from a dynamical screening of the
Coulomb potential remaining in the frame of a single physical
mechanism for both supersonic and subsonic regimes. The
distinction of the dynamical screening from the static Debye
screening is due to a motion of particles relative to screening
charges (counter-ions or electrons). From the physical point of
view, this effect can be interpreted as a consequence of loss of
spherical symmetry of the Debye screening cloud around the moving
charged particle resulted in a space charge with the opposite sign
forming in its wake. Not only do this space charge compensates the
particle's potential but may also give rise to a local wake
potential of the opposite sign. In the supersonic regime the wake
potential oscillates inasmach of Cherenkov wave generation.

This effect is well known in the usual electron-ion plasma
\cite{11,12,16}. When the charge is moving, the static Debye
screening modifies so that the potential in the oncoming flow of
electrons grows approaching to the Coulomb potential with an
increasing of velocity while the potential in the outward flow
decreases up to alternating of its sign.

Similar effect was found \cite{16,14} in a system of gravitating
masses where the static screening is absent and accounting for the
dynamical screening gives rise to an alternating potential of
gravitational interaction.

The aim of the present paper is to propose the most general model
of dynamical screening of field of dusty particle charge to
explain the observable interparticle attraction. It may be
supposed that the distance from the particle to the attractive
minimum of the wake potential determines a period of the resulted
lattice of dusty particles.

Besides, the interparticle attraction is to give rise to a surface
tension of a dusty phase interface. Really, this explains a sharp
non-diffusion character of a particle density variation observed
at the interfaces in the experiments on dusty plasma crystals. The
most characteristic phenomenon of such kind are voids in a
homogeneous dusty plasma. The characteristic lenticular form of
the voids may be explained by strong peculiar dependence of the
surface tension on the surface orientation relative to the ion
flow.

The notion of the surface tension of a dusty phase was mentioned
briefly by Tsitovich \cite{tsit1} but his estimation of its value
was erroneous. Conceivably that might be the reason why the
surface tension was not mentioned in the subsequent papers by
Tsitovich (as a co-author) on the theory \cite{tsit2,tsit3} of
spherical voids.

The outline of the paper is the following. In Sec. II, the model
for effective non-potential attractive forces of interactions
between dust particles will be developed. This model can explain
an appearance of the lattice structure both in gas discharge dusty
plasma and in thermal dusty plasma of gas burner. In Sec. III, a
surface tension of a boundary of dusty phase is estimated and
conditions of stability of a lenticular void in the dusty plasma
are examined.

\section{Effective interparticle interaction}

The system under consideration consists of negatively charged
dusty particles of the charge $Q=-Z_pe$ and concentration $n_p$,
and positively charged counter-ions of the charge $e$ (for
definiteness sake we will consider singly charged ions) and mass
$m$. For simplicity, neutral molecules of a buffer gas are
neglected here and dusty particles are considered as point
charges.

While forming of the negative charges of dusty particles the
electrons are condensed on particles' surfaces and their
concentration in plasma decreases sufficiently. As a result, in
the case of great $Z_pn_p$ their density becomes so little that
the electron Debye radius of screening of the ion charge appears
to be much greater the ion Debye radius of screening. Thus, we can
restrict our consideration to the simplified model of
one-component ion plasma in which the negatively charged dusty
particles are immersed.

If the negative test point charge $Q$ is moving in the system of
ions and dusty particles with the velocity ${\bf u}$ it gives rise
to some perturbations of the system state. Due to great difference
in masses and concentrations of ions and dusty particles we can
restrict our consideration to a perturbation of the ion component
only (just as a perturbation of an electron component is taken
into account only, as a rule, in the problem of screening of an
ion charge in the electron-ion plasma). Such perturbation of ionic
subsystem is described by the set of the Vlasov equation for a
distribution function over coordinates and velocities of ions and
Poisson equation for an effective potential induced by the
perturbation of ion density and moving test charge. Interactions
between ions and neutral molecules of the buffer gas are neglected
here \cite{ther}.

Such simplified model is well studied in the test-particle
approach to the electron-ion plasma theory \cite{11,12,16} so we
can omit intermediate calculations and write down at once the
resulted form of the effective potential of the moving test charge
\begin{equation}
\Phi ({\bf r},t)=\frac Q{2\pi ^{2}}\int d^3 k \frac {\exp\{{\bf
k\cdot (r}-{\bf u}t)\}}{k^2 \varepsilon ({k},{\bf k\cdot u})}
\end{equation}
where
\begin{equation}
\varepsilon ({k},{\bf k\cdot u})=1+\frac{\kappa ^{2}}{k^{2}}
W\left(\frac{{\bf k\cdot u}}{k\tilde{v}}\right)
\end{equation}
is the dynamical permittivity of the ion subsystem. Here  $\kappa
=\left( 4\pi e^{2}n_{i}/(k_BT_i)\right) ^{1/2}$ is the ionic Debye
wave number, $\tilde{v}= (2k_B T_i/m)^{1/2}$ is the mean heat
velocity of the ions and
\begin{equation}
W(t)=1-\sqrt {\pi}te^{-t^{2}}\,{\rm erfi}(t)+ i\sqrt
{\pi}te^{-t^{2}}.
\end{equation}

\begin{figure}
\includegraphics[scale=.5]{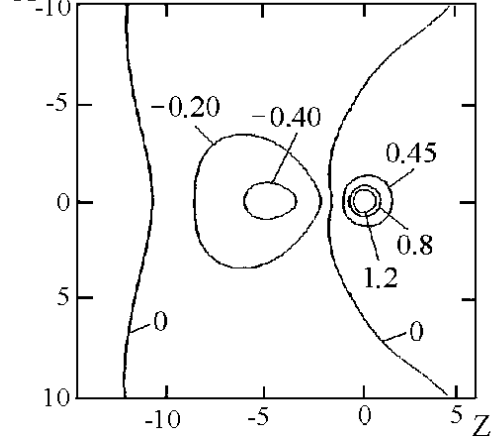}
\caption{The reduced potential $\varphi = \Phi /(Q\kappa)$ of the
negatively charged dusty particle at $Z=0$ moving relative ions
along axis $Z$ with the reduced velocity $M=u/\tilde v=3$.}
\end{figure}

Let us choose an axis $z$ along ${\bf u}$ and introduce
dimensionless variables $Z=(z-ut)\kappa,\,\,X=x\kappa,\,\, {\bf
K}={\bf k}/\kappa $ and the Mach number $M=u/\tilde{v}$ relative
to the ion heat velocity. Then, in the test particle's frame
accounting for the cylindrical symmetry, we get
\begin{widetext}
\begin{equation}
\Phi ({\bf r}) =\frac {Q\kappa}{2\pi^2 }\int^\pi_{-\pi}d\phi
\int^\pi_0d\theta\int dK
K^2\sin\theta\frac{\exp\{iK\Delta\}}{K^2+W(M\cos\theta)}=
Q\kappa\varphi(X,Z;M),
\end{equation}
$$\varphi (X,Z;M)=\frac 1{(X^2+Z^2)^{\frac 1{2}}}- I(X,Z;M), $$
where
\begin{equation}
I(X,Z;M)=\frac {1}{2\pi^2 }\int^\pi_{-\pi}d\phi
\int^\pi_0d\theta\int dK\,W(M\cos\theta)\sin\theta\frac{\exp
\{iK\Delta\}}{K^2+W(M\cos\theta)},
\end{equation}
$$\Delta = X\sin\theta\cos\phi+ Z\cos\theta. $$ This integral
determines the departure of the effective potential from the
Coulomb potential to which the first member of $\varphi (X,Z;M)$
corresponds. When the Cauchy integral over $K$ is taken, we find
\begin{eqnarray}
I(X,Z;M) &\dh=&\dh\frac {1}{4\pi^2}\int^\pi_{-\pi}d\phi
\int^\pi_0d\theta \sin\theta \sqrt{W}\, \Bigl[2\,i\,\sinh
({\Delta}{\sqrt{- W}}){\rm Ci} (-i\,{\Delta}{\sqrt{W}})+
\Bigr.\nonumber\\ &\dh {}&\dh + \Bigl.\cosh ({\Delta}{\sqrt{W}})
\,\left(\pi + 2\,i \,{\rm Shi}({\Delta} {\sqrt{W}})\right)\Bigr].
\end{eqnarray}
\end{widetext}
It is not difficult to calculate numerically the
value of this integral in an arbitrary range of the parameter $M$
variations at any point $X,\,Z$.

Corresponding results of numerical calculations for $M=3$ are
illustrated in Fig.1. It is seen here that there is well defined
negative minimum of the potential along the axis $Z$ at a distance
of order of five Debye radius, which has been formed as a result
of deformation of screening Debye cloud in the vicinity of the
moving particle.

In gas discharges, where forming of crystal structure of dusty
particles was observed, a diffusion ion velocity $\bf u$ is
determined by their mobility and an electric field strength. To
define the mobility, we are to take into account a tendency of
ions to unite with molecules and atoms into complexes of the types
$N_4^+,\,O_4^+,\,Ne_2^+,\,He_2^+,$, and recharge effect also.
According to \cite{19}, the diffusion velocity of the complex
$Ne_2^+$ is of order of $50 \,{\rm m\,c^{-1}}$ in a characteristic
for glow-discharge field $E/p=1\,{\rm V\,cm^{-1}torr^{-1}}$ and
the heat velocity $\tilde v=400\,{\rm m\,c^{-1}}$ (at $T=300\,K$).
Then $M=0.125$. For the ion $Ne^+$, the diffusion velocity
decreases near-threefold as a consequence of the resonant
recharge, then $M\simeq 0.05$. In the case of strong field, the
ion drift velocity may not only approach to the heat velocity but
surpass it \cite{19}. Because of this, a wide range of values of
the parameter $M=u/\tilde v$ will be considered below (asymptotic
case $M\gg 1$ was discussed in \cite{20,21}).

Observed quasi-crystal structure in a thermal plasma, that is, in
the dusty flame of gas burner \cite{6,10,22}, can be treated also
as a consequence of movement of charged particles relative to ions
of the flame. To obtain quantitative estimations of the effect,
the additional data on a gap of velocities of particle and ion
flows are necessary. This gap is determined by conditions of
injection of the dusty particles into the flame, and in the steady
state limit it is likely to go to a sedimentation velocity of
charged particles in the rising flow.

\begin{figure}[b]
\includegraphics{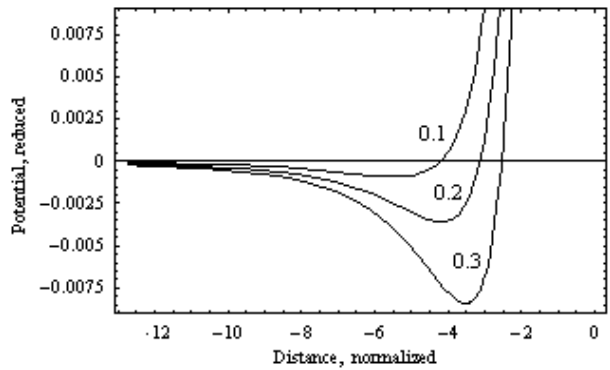}
\caption{The same as in Fig. 1 along the axis $Z$ at
$M=0.1,\,\,0.2,\,\,0.3$, correspondingly.}
\end{figure}

The graphs of $\varphi (X=0,Z<0;M)$ for different $M$ in the wake
of charged particle are illustrated in Figs. 2,3. As can be seen,
the depth of the negative minimum $|\varphi_{min}|$ increases with
an increasing of $M$ from 0.01 to 1  (see Fig.4), and its location
$Z_{min}$ shifts to the particle. At low $M$ $|Z_{min}|$ depends
on $M$ as $|Z_{min}|=1.6-1.7\ln M$ (see Fig.5). When further
increasing of $M$, the potential becomes oscillating and its first
minimum moves away from the particle. For $M>2$ the graph in Fig.
5 can be well approximated by the expression $|Z_{min}|=
1.8\sqrt{M^2 -1}$, which is close to the experimental data
\cite{1,2,3,4} on the period $L$ of a crystal lattice in a dusty
plasma if it is identified as $L\simeq |Z_{min}|$. It should be
noted here that experimental detection of the minimum of $L(M)$
corresponding to the minimum of the graph in Fig. 5 would be a
good evidence in favor of the approach under discussion.

The values $M>1$ correspond to a supersonic movement when
Cherenkov radiation of ion-acoustic waves takes a place that
explains an appearance of space oscillations of the wake
potential. It should be noted that in the range of values of $M$
from 1 to $\sim 3.3$ an increasing of the depth of the first
minimum takes a place but at further gain of $M$ the depth of the
first minimum decreases (see Fig.4).
\begin{figure}[b]
\includegraphics[scale=1.09]{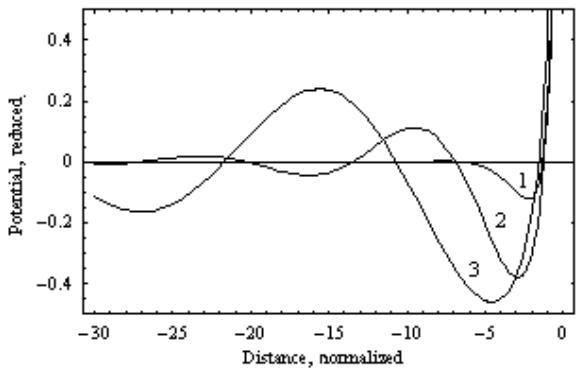}
\caption{The same as in Fig. 1 along the axis $Z$ at
$M=1,\,\,2,\,\,3$, correspondingly.}
\end{figure}

In the incoming flow ($Z>0$) the effective potential varies
continuously with a gain of $M$ from the Debye potential
$\sim\exp\{-Z\}/z$ to Coulomb one $\sim 1/z$. This indicates that
the forward part of the Debye screening cloud is not pressed by
the incoming flow to the particle but blew away from it.

The negative minimum of the wake potential points to presence of a
space charge of the opposite sign. Thus, the large charged
particles moving relative ion subsystem together attendant space
charges constitute a system of dipoles oriented along the
direction of the relative motion. This gives rise to formation of
hierarchies of particles-dipoles along lines of the ion flow.

Moreover, this interaction between dust particles is asymmetric in
such a way that attractive force is communicated only downstream
the ion flow. This situation was clearly demonstrated in the
experiments \cite{tak,melt2} where upper or lower particle was
pushed by the laser beam and it was just lower dust particle that
fitted its position when the upper particle was shifted and not
vice versa. So heavily non-conservative character of the
interparticle interactions excludes any possibility to introduce
an effective potential of the interaction.

\begin{figure}
\includegraphics[scale=1.17]{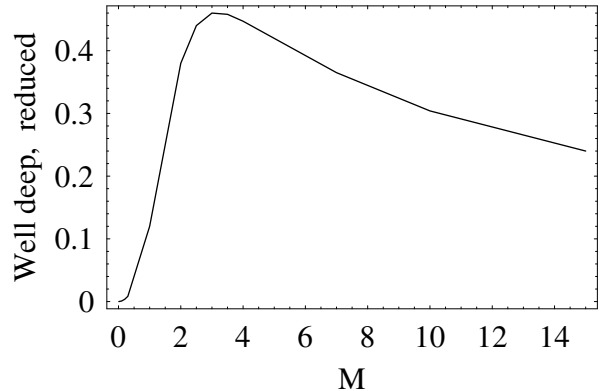}
\caption{Dependence of the depth of negative minimum of the
reduced potential on $M$.}
\end{figure}

Besides, as it is seen in Fig.1, the resulted negative potential
is long-range in the transverse direction $X$, also, that provides
an attraction between particles from neighbor hierarchies and, in
its turn, causes the particles from neighbor hierarchies to shift
relative one another along the ion flow for a distance of order of
a half dipole length. Because of this interaction is much weaker
then along the flow, the situations can take place when the
lengthwise attraction appears quite strong for the hierarchies
formation but too weak for transverse ordering \cite{10}.

Thus, the movement of the charged particle relative ion flow gives
rise to anisotropic potential field in which an energy of another
similar particle of the charge $Q$ is $U({\bf r})= Q\Phi ({\bf
r})=Q^2\kappa \varphi ({\bf r})$. Then, the effective force of
interaction between the particles is
\begin{equation}
{\bf F} ({\bf r})=- \frac{\partial U({\bf r})}{\partial {\bf
r}}=-Q^2\kappa\frac{\partial \varphi({\bf r})}{\partial {\bf r}}
\end{equation}

For the formation of the crystal structure, it is necessary that
the maximum deep of the attraction energy $|U_{min}|=|Q\Phi_{min}|
= Q^2\kappa|\varphi_{min}|$ surpass the heat motion energy, that
is,
\begin{equation}
\frac {|U_{min}|}{k_BT_p}>1,\,\,\,{\rm or}
\,\,\,|\varphi_{min}|>\frac {k_BT_p}{Q^2\kappa}
\end{equation}
where $T_p$ is the kinetic temperature of the particle. According
both experimental data \cite{23} and theoretical estimations
\cite{24} $T_p$ may sufficiently surpass the ion temperature. For
definiteness sake, we take $T_p=1000\,Ê$, $Z_p=10^4,\,\,
\kappa=400\,cm^{-1}$, then the condition of the hierarchy
stability becomes $|\varphi_{min}|>10^{-5}$. According to the
approach under discussion (see the graph in Fig. 4), this
condition is fulfilled when $M>0.01$.
\begin{figure}
\includegraphics[scale=.88]{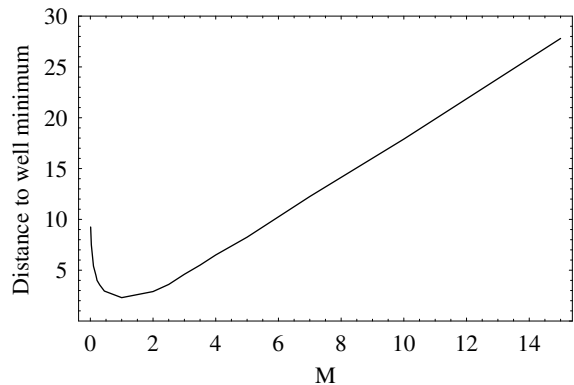}
\caption{Dependence of the position of negative minimum of the
reduced potential on $M$. }
\end{figure}

The long-range attraction between likely charged dusty particles
resting in an equilibrium plasma ($M=0$) was found in  \cite{25}.
It is possible that this kind of attraction can give rise to
forming of crystal structure but it is irrelevant to the observed
\cite{1,2,3,4} dependence of the lattice period on the ion flow
velocity and forming (at small $M$) of hierarchies along the flow
non-interacting with which other.

\section{Surface tension of dusty phase}

One of characteristic feature of a dispersed phase in dusty plasma
is presence of sharp (non-diffusion) boundary surface of a dusty
cloud ("dusty drop"). Considerable recent attention has been
focused on the problem of formation of voids ("dusty bubbles") of
the dispersed phase. It is reasonable to suppose that both
phenomena are of the same nature related with presence of a
surface tension resulted from the attractive interaction between
charged particles.

In general, the surface tension coefficient of an interphase
boundary is defined as the integral of a difference between
transverse $p_\bot$ and longwise $p_{\|}$ pressures over an
interphase layer of thickness $l$
\begin{equation}
\gamma =\int_{-l/2}^{l/2}(p_\bot(z_1) - p_{\|}(z_1))dz_1
\end{equation}
where the axis $z_1$ directed along the normal to the layer.

On the molecular level, a pressure can be written in the form of a
virial equation of state and expressed via a number density $n(z
_1)$, virial of a force of intermolecular interaction ${\bf
F}({\bf r})$ and radial correlation function $g({\bf r})$, where
${\bf r}={\bf r}_{2}-{\bf r}_{1}$ is the distance between
particles. For a bulk equilibrium fluid the function $g({\bf r})$
is calculated as a rule with the use of the Born--Green equation.
At an interphase surface the situation is much more complex; this
being so, notably rough approximations are used here. One of them
is Fowler's step approximation, when the number density $n(z_1)$
is regarded as constant $n_0$ within the liquid phase and zero
outside the liquid, and radial function $g({\bf r})$ within the
surface layer is assumed to be the same as in the bulk liquid.
Then the surface tension coefficient is determined by the Fowler
formula \cite{26,27}
\begin{equation}
\gamma =-\frac{\pi n_0^2}{8}\int_0^\infty r^4 F_r(r)g(r)
dr,\,\,\,z>0.
\end{equation}
It is implied here that the $z$-component of the vector ${{\bf
r}}$ is normal to the surface. Kirkwood and Buff \cite{28}
estimated the surface tension of liquid argon on the base of this
formula and obtained quite satisfactory value (error $\sim 25\%$).

It is supposed in the Fowler' formula that the intermolecular
interaction is spherically symmetric that does not allow to apply
this formula as such to the boundary of dusty phase.

\begin{figure}
\includegraphics[scale=0.615]{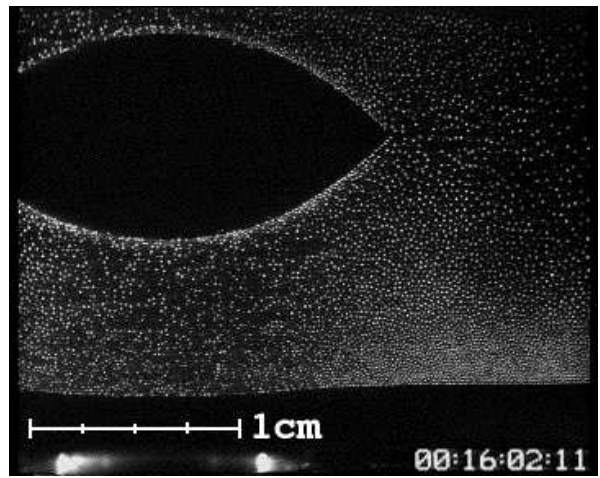}
\caption{The typical void in dusty plasma. (Copied from the Web
page
http://www.mpe.mpg.de/www\_th/plasma-crystal/PKE/PKE-Nefedov.html
dedicated to the memory of A. P. Nefedov. Authorized by Max Planck
Institute of Extraterrestrial Physics (Germany))}
\end{figure}

Rederiving the Fowler formula with due regard for the anisotropy
of interparticle interaction we obtain
\begin{equation}
\gamma =-\frac{n_0^2}{2}\int \left[z^2F_z({\bf r})- xzF_x({\bf
r})\right]g({\bf r}) d{\bf r},\,\,\,\,\,z>0.
\end{equation}
where $F_z$ and $F_x$ are $z$ and $x$ components of the force (7).

When applying to the boundary of dusty phase, the pressures
$p_\bot$ and $p_{\|}$ are to be considered as partial pressures of
dusty component. Then, the surface tension of dusty phase should
be interpreted as partial property too.

To estimate its value some simplification will be introduced.
Firstly, the radial distribution will be used in the form of the
Boltzmann factor $$g({\bf r})= \exp\{-\frac{U({\bf r})}{k_BT}\},$$
that is typical for dilute systems. Secondly, when estimating the
surface tension, the interactions within hierarchies will be
regarded only as it is much stronger interactions between
particles from different hierarchies. Then, for surfaces oriented
across and along the ion flow we get
\begin{equation}
\gamma_\bot =-\frac{n_p^2}{2}\int z^2F_z({\bf r}) g({\bf r}) d{\bf
r},\,\,\,\,\,z>0.
\end{equation}
\begin{equation}
\gamma_{\|} =\frac{n_p^2}{2}\int  xz F_x({\bf r})g({\bf r}) d{\bf
r},\,\,\,\,\,z>0.
\end{equation}
The axis $z$ in both equations directed along the normal to the
surface and the forces $F_z({\bf r})$ in the equation for
$\gamma_\bot$ and $F_x({\bf r})$ in the equation for $\gamma_{\|}$
are determined by interactions along hierarchies.

A numerical estimation of these expressions for $n_p\sim
10^6\,{\rm cm^{-3}}$ and $M=0.01$ leads to $\gamma_\bot\sim
10^5\,{\rm din\,cm^{-1}}$ and $\gamma_{\|}\sim -10^4\,{\rm
din\,cm^{-1}}$. Both of these coefficients in absolute magnitude
surpass sufficiently the typical surface tension coefficient of a
liquid at normal conditions $\gamma_{liq}\simeq 10^2\,{\rm
din\,cm^{-1}}$ (as an example, for water $\gamma_{liq}=70\,{\rm
din\,cm^{-1}}$). Such great difference is quite natural. Really,
the negative minimum $U_m$of the potential $U$ at given conditions
is of order of $-10^{-12}\,{\rm erg}$ while the depth of the
potential well of, for example, the Lennard-Jones potential for
noble gases is of order of $-10^{-14}\,{\rm erg}$. Inasmuch as the
potential $U$ enters (with sign minus) into the exponent of the
radial distribution function, the surface tension coefficient
appears to be very sensitive to its value. As a result, we have
$|\gamma_{\bot , \,\|}| \gg\gamma_{liq}$ in spite of
$\gamma\propto n^2$ and $n_p\ll n_{liq}$.

Hydrodynamic stability of a spherical gas bubble or liquid drop is
determined by value of a Weber number $We=\rho v^2 a/\gamma$,
which is the ratio of a dynamic head $\rho v^2$ to a surface
tension pressure $\gamma/a$, where $a$ is the radius of the bubble
or drop. For the case of the void in dusty plasma, the dynamic
head $\rho v^2$ is determined by the flow of ions and neutral
gases with the velocity $v=u = \tilde v\,M \simeq 5\cdot
10^4\,M\,{\rm cm\,c^{-1}}$. We can put $\gamma\simeq \gamma_\bot$
for the slightly curved surface of a lenticular void with the
radius of curvature $a\simeq 2\,{\rm cm}$ sufficiently greater its
size. Then at characteristic pressure of order of $0.5 {\rm torr}$
we have $We\simeq 10^4\,M^2/\gamma_\bot$. The coefficient
$\gamma_\bot$ increases with increasing of $M$, so that the Weber
number $We$ attains its the most great value at the minimal from
considered here values of $M$, i.e. at $M=0.01$ when
$\gamma_\bot\simeq 10^{5}$. Then $We_{max}\simeq 10^{-4}$. When
the Weber number is so small the surfaces of the lenticular void
are stable, certainly. Their deviations from a spherical form can
be due to variations of $\gamma$ relative to $\gamma_\bot$ as a
consequence of inevitable (although small) distortions of the
condition of orthogonality to the flow on the curved surface.

Stability of the lenticular void ensures possibility of its
existence. The problem of its appearance remains to be solved. It
may be suggested that a phase transition of the first kind takes a
place in a homogeneous dusty plasma, and strong anisotropic
surface tension can give rise to a peculiar nucleation process.
\section{Conclusions and discussions}
Attractive interactions between likely charged colloidal or dust
particles in plasma have been discussed in this effort in a view
to explain the observed crystal structure formed from these
particles. The crucial point of the model under discussion is the
presence of the counter-ion flow. Its velocity $u$ relative the
dust particles is not necessary to exceed the heat velocity
$\tilde v$ of counter-ions (or ion--acoustic velocity) as it was
supposed in ref. \cite{20}. Even at small $u$ when $u/\tilde
v\simeq 10^{-2}$, the effective attractive interaction appears to
be sufficiently strong to ensure stability of regular hierarchies
of particles oriented along the ion flow.

Another consequence of strong attractive interparticle
interactions is the partial surface tension $\gamma $ of the dusty
phase boundary.

This concept was briefly discussed by Tsytovich \cite{tsit1} who
estimated $\gamma$ as a work necessary for construction of a bulk
liquid of the unit surface and height $h$. As a result, he
obtained $\gamma=U_m n_p h$, then he took $|U_m|=100\,{\rm eV}$
that correspond to our result at $M=1$ and found $\gamma \simeq
10^{-2}{\rm din\,cm^{-1}}$ neglecting the negative sign of $U_m$.
If he accounted for $U_m<0$ he would get $\gamma <0$, that is
natural as the work for construction of the coupled state is to be
negative.

Here, the surface tension is estimated on the base of a
generalization of the Fowler's formulae taking into account the
anisotropy of interparticle interactions. As a result, the strong
dependence of $\gamma $ on orientation of the dusty interface is
found. In particular, when $u/\tilde v \simeq 10^{-2}$ we get
$\gamma_{\bot }\simeq 10^5\,{\rm din\,cm^{-1}}$ and
$\gamma_{\|}\simeq -10^4\,{\rm din\,cm^{-1}}$ for surfaces
oriented across and along the ion flow, correspondingly. So great
positive and negative value results in the characteristic
lenticular form of the voids in dusty plasma having no surfaces
along the ion flow. Such a form was observed both under
micro-gravity condition (see Fig. 6) and in terrestrial
experiments by Samsonov and Goree (see Fig. 5d in Ref.\cite{sam}).
They observed also an appearance of a void mode as a penetration
of a finger-shaped (in a vertical section) dusty free region
through the side boundary of the gas discharge (see Fig. 5c in
Ref.\cite{sam}). A sharp end of the finger and its fast travel
across the gas discharge volume count in favor of negative surface
tension of the interface at its end. It means that an account for
non-spherical form of the void is necessary for any theoretical
model of the void.
\subsection*{Acknowledgements} I acknowledge
useful discussions with A.V. Vityazev and a discussion of the void
problem with S.I. Popel.


\begin{thebibliography}{99}
\bibitem{1} J. H. Chu and I. Lin, Phys. Rev. Lett. {\bf 72}, 4009 (1994).

\bibitem{2} H. Thomas, G. E.
Morfill, V. Demmel et al., Phys. Rev. Lett. {\bf 73}, 652 (1994).

\bibitem{3} Y. Hyashi and K. Tachibana, Jap. J. Applied Phys. {\bf
33}, L804 (1994).

\bibitem{4} A. Melzer, T. Trottenberg and A.
Piel, Phys. Lett. A {\bf 191}, 301 (1994).

\bibitem{6} V. E. Fortov, A. P. Nefedov, O. F. Petrov
et al. JETP {\bf 84}, 256 (1997).


\bibitem{8} A. Barkan and R. L. Merlino, Phys. Plasmas {\bf
2}, 3261 (1995).

\bibitem{9} V. E. Fortov, A. P. Nefedov, V. M.
Torchinky, Phys. Lett. A {\bf 229}, 317 (1997).

\bibitem{10} A. P.
Nefedov, O. F. Petrov, V. E. Fortov, Usp. Fiz. Nauk, {\bf 167},
1215 (1997) [Phys. Usp. {\bf 40}, 1215 (1997)].

\bibitem{17} O. M. Belotzerkovskii, I. E. Zakharov, A. P. Nefedov et al., JETP {\bf
88}, 449 (1999).

\bibitem{20} S. V. Vladimirov and M. Nambu, Phys. Rev. E
{\bf 52}, R2172 (1995).

\bibitem{21} F. Melands\o and J. Goree,
Phys. Rev. E {\bf 52}, 5312 (1995).


\bibitem{vlad} S. Benkadda, V. N. Tsitovich and S. V. Vladimirov, Phys.
Rev. E {\bf 60}, 4708 (1999).


\bibitem{11} P. Chenevier, J. M. Dolique and H. Peres, J.
Plasma Phys. {\bf 10}, 185 (1973).

\bibitem{12} T. Peter, J. Plasma Phys. {\bf 44}, part 2, 269
(1990).

\bibitem{16} A. G. Bashkirov and A. V. Vityazev,
Physica A {\bf 305}, 271 (2002).

\bibitem{14} A. G. Bashkirov and A. V. Vityazev,
Astroph. Journ. {\bf 457}, Part 1, 10 (1998).

\bibitem{tsit1} V. N. Tsitovich, Usp. Fiz. Nauk, {\bf 167}, 57 (1997) [Phys. Usp. {\bf 40},
53 (1997)].

\bibitem{tsit2} J. Goree, G. E. Morfill, V. N. Tsitovich and S. V. Vladimirov, Phys.
Rev. E {\bf 59}, 7055 (1999).

\bibitem{tsit3}  V. N. Tsitovich, S. V. Vladimirov, G. E. Morfill and J. Goree, Phys.
Rev. E {\bf 63}, 056609 (2001).

\bibitem{ther} There is no problems to account them by the way of adding a
ion-neutral collision integral to the right-hand-side of the
Vlasov equation as it was done, for example, in \cite{16,18} where
conditions of forming of quasi-crystal structure in charged
colloid solute were found.

\bibitem{18} A. G. Bashkirov, Russian J. Phys.
Chemistry {\bf 74}, Suppl. 1, S59 (2000).

\bibitem{19} Yu. P. Rayzer, {\it Fizika gasovogo razryada (Physics of gas discharge)}, Nauka, Moscow
(1992).

\bibitem{22} A. P. Nefedov, O. F.
Petrov and V. E. Fortov, Russian J. Phys. Chemistry {\bf 74},
Suppl. 1, S136 (2000).


\bibitem{tak} K. Takahasi, T. Oishi, K. Shimomai et al. Phys.
Rev. E {\bf 58}, 7805 (1998).

\bibitem{melt2}  A. Melzer, V. A. Schweigert and A. Piel, Phys.
Rev. Lett. {\bf 83}, 3194 (1999).


\bibitem{23} V. V. Zhakhovskii, V. I.
Molotkov, A. P. Nefedov et al. Letters JETP {\bf 66}, 419 (1997).

\bibitem{24} A. G. Zagorodny, P. P. G. M. Schram and S. A.
Trigger, Phys. Rev. Lett. {\bf 84}, 3594 (2000).

\bibitem{25} A. S. Ivanov,  Phys. Lett. A {\bf 290}, 304 (2001).

\bibitem{26} S. Ono, S. Kondo, Handbuch der Phys., {\bf 10}, 134
(1960).

\bibitem{27} J. S. Rowlinson, B. Widom, Molecular theory
of capillarity. Clarendon Press. Oxford. 1982.

\bibitem{28} J. G. Kirkwood and F. P. Buff, J. Chem. Phys. {\bf 17}, 338 (1949).

\bibitem{sam} D. Samsonov and J. Goree, Phys.
Rev. E {\bf 59}, 1047 (1999).

\end{thebibliography}
\end{document}